\documentclass[aps,preprint,epsfig,rotate]{revtex4}
\usepackage{graphicx}
\usepackage{bm}
\usepackage{epsfig}

\input epsf
\begin{document}     
       
\title{On the general theory of bound state spectra in the Coulomb few- and many-body systems}

\author{Alexei M. Frolov}
\email[E--mail address: ]{alex1975frol@gmail.com ; afrolov@uwo.ca}


\affiliation{Department of Applied Mathematics \\
       University of Western Ontario, London, Ontario N6H 5B7, Canada} 

\date{\today}

\begin{abstract}
Based on the fact that the Hamiltonians of the Coulomb many-particle systems are always factorized we develop the two different approaches for analytical 
solution of the Schr\"{o}dinger equation written for arbitrary few- and many-particle Coulomb systems. The first approach is the matrix factorization method. 
Another method is based on the $D^{+}-$series of representations of the hyper-radial O(2,1)-algebra. The both these methods allow us to obtain the closed 
analytical formulas for the bound state energies in an arbitrary many-particle Coulomb system. \\

\noindent
This manuscript has been accepted for publication in Chem. Phys. Lett. \\

\noindent 
PACS number(s): 31.15.-A, 31.15.ac and 32.30.-r
\end{abstract}

\maketitle
\newpage

\section{Introduction}

In this communication we discuss a few diferent approaches which can be used for analytical determination of the bound state spectra in actual atoms and ions, 
i.e. in one-center Coulomb systems which contains a number of bound electrons $N_e (\ge 1)$. To investigate such systems in our previous study \cite{Fro1} we have 
developed the method of matrix factorization which can be applied to the Hamiltonians of arbitrary many-particle Coulomb systems, including many-electron atomic 
Hamiltonians. This method is based on the fact \cite{Fro1} that the Hamiltonian of an arbitrary many-particle system is always factorized, i.e. it is represented 
in the form of a product of the two differential operators of the first-order in respect to the non-compact variable(s). This approach is very effective in 
hyperspherical coordinates \cite{Fock}, \cite{Knirk} where we have only one (unique) non-compact variable - the hyper-radius $r$ of the whole system \cite{Fro1}. 
The matrix factorization method allows one to determine the bound state spectra of arbitrary few- and many-electron atoms and ions. Moreover, all analytical formulas 
derived in this method are relatively simple and threir derivation is physically transparent. Formally, to obtain such formulas one needs to solve a few simple matrix 
equations for three infinite-dimensional matrices. 

Unfortunately, all attempts to generalize the method of matrix factorization to other sets of coordinates used in atomic and molecular physics were never successful. 
Very likely, we cannot develop similar methods for other sets of coordinates which are used to describe one-center few- and many-electron Coulomb systems and include 
two or more non-compact variables. Futher investigation of this problem based on arguments presented in \cite{Fock} and \cite{Fock1} brought us to the conclusion 
that such a very special role of hyperspherical coordinates is closely related to the known fact that in hyperspherical coordinates the discerete and continuous parts 
of the energy spectrum in arbitrary Coulomb many-body systems (including atoms and ions) are completely separatred from each other with the use of a simple algebraic 
(even arithmetic) transformation of the wave function. Furthermore, such an unitary transformation is written as an operator which depends upon the hyper-radius
$r$ only. For other sets of coordinates such a separation of the two parts of energy spectrum is not possible to perform. 

It is shown in this study, such a separarion of the discrete and continuous parts of energy spectra in an arbitrary many-particle Coulomb system is closely related 
with the existence of a three-operator, hyper-radial O(2,1)-algebra. This non-compact algebra can be constructed for arbitrary Coulomb few- and many-body 
(non-relativistic) systems, including various atoms, ions, clusters and molecules. Based on this hyper-radial O(2,1)-algebra and some of its representations we have 
developed another effective method which can be used to determine analytical solutions of the Schr\"{o}dinger equations written for different Coulomb few- and 
many-body systems.    

In this study we apply the hyperspherical coordinates introduced in \cite{Fock}, \cite{Knirk}. In hyperspherical coordinates (definitions of these coordinates and 
thier properties can be found, e.g., in \cite{Fock}, \cite{Knirk} and \cite{Smirn}) the corresponding time-independent Schr\"{o}dinger equation takes the form $H(r, 
\Omega) \Phi(r, \Omega) = E \Phi(r, \Omega)$, where $\Phi(r, \Omega)$ is the unknown wave function of the atomic system, $E$ is the `eigenvalue' of the operator 
$H(r, \Omega)$ is the Hamiltonian which is written in the following form \cite{Fock}, \cite{Knirk}
\begin{eqnarray}
  H(r, \Omega) =  -\frac{1}{2} \Bigl[ \frac{\partial^2}{\partial r^2} + \frac{3 N_e - 1}{r} \frac{\partial}{\partial r} - \frac{\Lambda^2_{N_e}(\Omega)}{r^{2}} 
 \Bigr]  + \frac{W(\Omega)}{r} \; \; \; \label{HHH}
\end{eqnarray}
where $\Lambda^2_{N_e}(\Omega)$ is the hypermomentum of the atom, while $W(\Omega)$ is the hyperangular part of the Coulomb interaction potential which includes
electron-nucleus and electron-electron parts and $N_e$ is the total number of bound electrons. Here and everywhere below in this study we apply the atomic units 
(where $\hbar = 1, \mid e \mid = 1$ and $m_e = 1$) and use the system of notation, which was defined in \cite{Fro1}. The operator $H(r, \Omega)$ Hamiltonian, 
Eq.(\ref{HHH}), is the Hamiltonian of the one-center many-electron (Coulomb) system, which is often called the atomic Hamiltonian. In particular, the notation $\Omega$ 
means the set of $(3 N_e - 1)$ angular and hyperangular electron's coordinates (compact variables), while $r$ designates the hyper-radius of the atom/ion. Note that 
$r$ is the unique non-compact variable of the problem and $0 \le r < +\infty$. 

By representing the wave functions $\Phi(r, \Omega)$ in the form $r^{-\frac{3 N_e - 1}{2}} \Psi(r, \Omega)$ one can reduce the operator $H(r, \Omega)$ (Hamiltonian) in 
Eq.(\ref{HHH}) to the following self-conjugate form
\begin{eqnarray}
  H(r, \Omega) =  -\frac{1}{2} \frac{\partial^2}{\partial r^2} + \frac{K^2_{N_e}(\Omega)}{2 r^{2}} + \frac{W(\Omega)}{r} \; \; \; \label{HH1}
\end{eqnarray}
which does not contain any linear derivative upon the hyper-radius $r$. The relation between the hyper-angular operators $\Lambda^2_{N_e}(\Omega)$ and $K^2_{N_e}(\Omega)$
is simple
\begin{eqnarray}
     K^2_{N_e}(\Omega) = \Lambda^2_{N_e}(\Omega) + \Bigl(\frac{3 N_e - 1}{2}\Bigr)^{2} - \Bigl(\frac{3 N_e - 1}{2} \Bigr)  \; \; \; \label{KH1} 
\end{eqnarray}
In other words, the difference between the hyperangular $K^2_{N_e}(\Omega)$ and $\Lambda^2_{N_e}(\Omega)$ operators is a numerical constant which depends upon the total 
number of particles $N_e$ in the Coulomb (or atomic) system. For atoms and ions the number $N_e$ coincides with the total number of bound electrons. The atomic Hamiltonian, 
$H(r, \Omega)$, written in the form of Eq.(\ref{HH1}), is called the self-conjugate form of atomic Hamiltonian, or self-conjugate atomic Hamiltonian, for short. Note that 
the both atomic Hamiltonians, Eqs.(\ref{HHH}) - (\ref{HH1}), contain one non-compact variable (the hyper-radius $r$) and multi-dimensional set of hyperangular and compact 
variables $\Omega$. In actual applications to analytical investigations it is very convenient to use one non-compact variable $r$. This can be achieved by representing all 
operators in the basis of hyperspherical harmonics. Then all operators take the mixed matrix-operator from, i.e. each of the matrix element is a differential operator upon 
the hyper-radius $r$. In the basis of the `physical' hyperspherical harmonics \cite{Fro1} ${\cal Y}_{\vec{K}(c),\vec{\ell}(c),\vec{m}(c)}(\Omega)$ (below HH, for short) and 
for the hyper-radial functions represented in the form $r^{-\frac{3 N_e - 1}{2}} \Psi(r)$ this Hamiltonian takes the following (self-conjugate) form
\begin{eqnarray}
  H(r) = \frac{1}{2} \Bigl[ p^{2}_r + \frac{\Bigl(\hat{K} + \frac{3 N_e + 1}{2}\Bigr) \Bigl(\hat{K} + \frac{3 N_e - 1}{2} - 1\Bigl)}{r^{2}} \Bigr] + \frac{\hat{W}}{r} 
  \; \; \; \label{HHHa}
\end{eqnarray}
where the hyper-radial momentum operator $p_r$ is defined as follows $p_r = (-\imath) \frac{\partial}{\partial r}$, $\hat{W}$ is the matrix of the hyperangular part of 
the Coulomb interaction potential in the basis of physical HH (definition of the physical HH can be found, e.g., in \cite{Fro1}, \cite{Our1} and references therein).  The 
commutation relation $[ p_r, r] = -\imath \hbar = -\imath$ is always obeyed for the $p_r$ operator and hyper-radius $r$. Also, in Eq.(\ref{HHHa}) the notation $\hat{K}$ 
stands for the matrix of hypermomentum which is a diagonal matrix in the basis of hyperspherical harmonics. The self-adjoint form of these operators is more appropriate 
for our present purposes. The matrices $\hat{W}, \hat{K}$ and a few other matrices arising below are assumed to be infinite-dimensional. In actual applications the 
dimensions of these matrices equal to the numbers of HH used. 

This paper has the following structure. In the next Section we re-consider the method of matrix factorization \cite{Fro1}. The principal difference between our current 
approach and method used in \cite{Fro1} is the use of all operators written in self-adjoint form. This allows us to correct some disadvantages of our earlier method 
\cite{Fro1}. In particular, the self-conjugate form of the momentum operator $p_{r}$ makes explicitly clear all relations between the both Heisenberg and Schr\"{o}dinger 
approaches in Quantum Mechanics. The O(2,1)-algebra of the three hyper-radial operators $S, T$ and $U$ is constructed and considered in the third Section. Based on the 
well-known theorem about self-adjoint representations of this algebra (see, e.g., \cite{Barut}, \cite{Perelom} and references therein) we develop an alternative method 
for analytical solution of actual few- and many-electron atomic problems. Concluding remarks can be found in the last Section. 

\section{Factorization of the Coulomb (atomic) Hamiltonians}

In this Section we apply the method of matrix factorization to the atoms and ions with $N_e-$bound electrons. Our goal is to show that our current representation of all 
operators in self-adjoint form leads to the same formulas for the energy spectrum of bound state, which were derived earlier in \cite{Fro1} with the use of these operators 
written in general form. In \cite{Fro1} (see also \cite{Fro1986}) we have shown that the atomic Hamiltonian $H(r)$, Eq.(\ref{HHHa}), is factorized, i.e. $H(r)$ can always 
be represented in the form
\begin{eqnarray}
  H = \Theta^{\ast}_1(r) \Theta_1(r) + \hat{a}_1 \; \; \; \label{factHa}
\end{eqnarray} 
where $\hat{a}_1$ is a matrix defined below, while the operator $\Theta_1(r)$ and its adjoint operator $\Theta^{\ast}_1(r)$ are the first-order differential operators defined 
in this study as follows 
\begin{equation}
 \Theta_1(r) = \frac{1}{\sqrt{2}} \Bigl[ -\imath p_r + \frac{\hat{\beta}_1}{r} + \hat{\alpha}_1 \Bigl] = 
 \frac{1}{\sqrt{2}} \Bigl[ -\frac{\partial}{\partial r} + \frac{\hat{\beta}_1}{r} + \hat{\alpha}_1 \Bigl] \; \; \; \label{Thm1}
\end{equation}
and  
\begin{equation}
 \Theta^{\ast}_1(r)  = \frac{1}{\sqrt{2}} \Bigl[ \imath p_r + \frac{\hat{\beta}_1}{r} + \hat{\alpha}_1 \Bigl]
 = \frac{1}{\sqrt{2}} \Bigl[ \frac{\partial}{\partial r} + \frac{\hat{\beta}_1}{r} + \hat{\alpha}_1 \Bigl] \; \; \; \label{Thm1c}
\end{equation}
where the notations $\hat{\beta}_1, \hat{\alpha}_1$ and $\hat{a}_1$ from Eq.(\ref{factHa}) stand for the symmetric, infinite-dimensional, in principle, matrices which do not 
commute with each other. The explicit forms of these operators upon variable $r$ are different from analogus operators defined in \cite{Fro1}. In actual applications the 
dimensions of these matrices coincide with the total number of hyperspherical harmonics used. By substituting these two expressions, Eqs.(\ref{Thm1}) - (\ref{Thm1c}), into 
Eq.(\ref{factHa}) one finds the following equations for the $\hat{\alpha}_1, \hat{\beta}_1$ and $\hat{a}_1$ matrices:
\begin{eqnarray}
    & & \hat{\beta}_1 (\hat{\beta}_1 - 1) = \Bigr(\hat{K} + \frac{3 N_e - 1}{2}\Bigl) \Bigr(\hat{K} + \frac{3 N_e - 1}{2} - 1\Bigl)  \; \; \; \label{eqx} \\
    & & \hat{\alpha}_1 \hat{\beta}_1 + \hat{\beta}_1 \hat{\alpha}_1 = 2 \hat{W} \; \; \; \label{eqy} \\
    & & \hat{a}_1 = -\frac12 \hat{\alpha}^2_1 \; \; \; \label{eqz}
\end{eqnarray}
where the matrix of hypermomentum $\hat{K}$ is a diagonal matrix in the basis of hyperspherical harmonics (or, in $K-$representation, for short). Solution of 
Eq.(\ref{eqx}) is written in the form 
\begin{equation}
    \hat{\beta}_1 = \hat{K} + \frac{3 N_e - 1}{2}   \; \; \; \label{eqxx}
\end{equation}
where we used the fact that the atomic wave functions of actual bound states must be regular at $r = 0$, i.e., at the atomic nucleus. As follows from this equation the matrix 
$\hat{\beta}_1$ is also diagonal in $K-$representation. Below, we apply only this $K-$representation, since it substantially simplifies a large number of formulas derived 
below. In particular, by using Eq.(\ref{eqxx}) and the formula from \cite{Belman} (see Chapter 10, \$ 18) we can write the explicit expression for the $\hat{\alpha}_1$ 
matrix
\begin{equation}
    \hat{\alpha}_1 = 2 \int_{0}^{+\infty} \exp(-\hat{\beta}_1 t) \hat{W} \exp(-\hat{\beta}_1 t) dt  \; \; \; \label{eqyy}
\end{equation}
Since the $\hat{\beta}_1$ matrix is diagonal, then for the $(ij)-$matrix element of the $\hat{\alpha}_1$ matrix one finds
\begin{equation}
  \Bigl[ \hat{\alpha}_1 \Bigr]_{ij} = \frac{2 W_{ij}}{[\beta_1]_{ii} + [\beta_1]_{jj}} = \frac{2 W_{ij}}{[\beta_{1}]_{i} + [\beta_{1}]_{j}} = 
  \frac{2 W_{ij}}{K_{i} + K_{j} + 3 N_e - 1}
\end{equation}
Finally, we can determine the $\hat{a}_1$ matrix from Eq.(\ref{eqz}). In particular, for the $(ij)-$matrix elements of the $\hat{a}_1$ matrix we obtain the formula 
\begin{equation}
  \Bigl[ \hat{a}_1 \Bigr]_{ij} = - 2 \sum_{k} \frac{W_{ik}}{\beta_{i} + \beta_{k}} \cdot \frac{W_{kj}}{\beta_{k} + \beta_{j}} = 
     - 2 \sum_{k} \frac{1}{\beta_{i} + \beta_{k}} \Bigl[ W_{ik} W_{kj} \Bigr] \frac{1}{\beta_{k} + \beta_{j}}  \; \; \; \label{eqzz}
\end{equation}

At the second stage of the procedure, we introduce the hyper-radial radial operators $\Theta_n(r)$ for $n = 2, 3, \ldots$, which are similar to the operators $\Theta_1(r)$ 
defined in Eq.(\ref{Thm1}), i.e.
\begin{equation}
 \Theta_n(r) = \frac{1}{\sqrt{2}} \Bigl[ -\imath p_r + \frac{\hat{\beta}_n}{r} + \hat{\alpha}_n \Bigl]
 = \frac{1}{\sqrt{2}} \Bigl[ - \frac{\partial}{\partial r} + \frac{\hat{\beta}_n}{r} + \hat{\alpha}_n \Bigl] \; \; \; \label{Thmn}
\end{equation}
The adjoint operators take the form 
\begin{equation}
 \Theta^{\ast}_n(r) = \frac{1}{\sqrt{2}} \Bigl[ \imath p_r + \frac{\hat{\beta}_n}{r} + \hat{\alpha}_n \Bigl]
 = \frac{1}{\sqrt{2}} \Bigl[ \frac{\partial}{\partial r} + \frac{\hat{\beta}_n}{r} + \hat{\alpha}_n \Bigl] \; \; \; \label{Thmnc}
\end{equation}
where $n = 2, 3, \ldots$. The logically closed method of matrix factorization is based on the following `ladder' conditions (or `ladder-like' equations, see, e.g., 
\cite{Green}) 
\begin{eqnarray}
  \Theta_n(r) \Theta^{\ast}_n(r) + \hat{a}_n = H_{n+1} = \Theta^{\ast}_{n+1}(r) \Theta_{n+1}(r) + \hat{a}_{n+1} \; \; \; \label{factHmn}
\end{eqnarray}
which must be obeyed for $n = 1, 2, \ldots$. By substituing the explicit expressions, Eqs.(\ref{Thmn}) and (\ref{Thmnc}) into Eq.(\ref{factHmn}) we obtain the following 
equations for the $\hat{\beta}_n, \hat{\beta}_{n+1}, \hat{\alpha}_{n}, \hat{\alpha}_{n+1}, \hat{a}_n$ and $\hat{a}_{n+1}$ matrices
\begin{eqnarray}
 & &  \hat{\beta}_{n+1} (\hat{\beta}_{n+1} - 1) = \hat{\beta}_{n} (\hat{\beta}_{n} + 1) \; \; \; , \; \; \; \label{I1} \\ 
 & & \hat{\alpha}_{n} \hat{\beta}_{n} + \hat{\beta}_{n} \hat{\alpha}_{n} = 2 \hat{W} = \alpha_{n+1} \hat{\beta}_{n+1} + \hat{\beta}_{n+1} \hat{\alpha}_{n+1} 
  \; \; \; , \; \; \; \label{I2} \\
 & & \hat{a}_{n} = -\frac12 \alpha^{2}_{n} \; \; \; , \; \; \; \hat{a}_{n+1} = -\frac12 \alpha^{2}_{n+1} \; \; \;   \label{I3}
\end{eqnarray} 
These matrix equations are similar to the analogous numerical equations derived in the traditional (or numerical) factorization method for the hydrogen-like atomic 
systems (see, e.g., \cite{Green}, \cite{Elut}). However, Eqs.(\ref{I1}) - (\ref{I3}) are written for the symmetric, infinite-dimensional matrices, which do not commute 
with each other, e.g., the $\hat{\beta}_n$ matrix do not commute with the $\hat{\alpha}_{n}$ and $\hat{a}_{n+1}$ matrices, etc. Solution of these equations, 
Eqs.(\ref{I1}) - (\ref{I3}), regular at $r = 0$ is written in the form
\begin{eqnarray}
 & & \hat{\beta}_{n+1} = \hat{\beta}_{n} + 1 = \ldots = \hat{\beta}_1 + n = \hat{K} + \frac{3 N_e - 1}{2} + n \; \; \; \label{eqxxm} \\
 & & \hat{\alpha}_{n+1} = 2 \int_{0}^{+\infty} \exp(-\hat{\beta}_{n+1} t) \hat{W} \exp(-\hat{\beta}_{n+1} t) dt  \; \; \; \label{eqyym} \\
 & & \hat{a}_{n+1} = -\frac12 \alpha^{2}_{n+1} \; \; \; \label{eqzzm}
\end{eqnarray}

From these equations one finds that Eq.(\ref{eqyym}) produces the explicit formula for the $(ij)-$matrix element of the $\hat{\alpha}_{n+1}$ matrix
\begin{equation}
  \Bigl[ \hat{\alpha}_{n+1} \Bigr]_{ij} = \frac{2 W_{ij}}{[\beta_{n+1}]_{ii} + [\beta_{n+1}]_{jj}} = \frac{2 W_{ij}}{[\beta_{1}]_{i} + [\beta_{1}]_{j} + 2 n} = 
  \frac{2 W_{ij}}{K_{i} + K_{j} + 3 N_e - 1 + 2 n}\; \;  \;
  \label{fres1}
\end{equation}
where $[\beta_{1}]_{i}$ is the $(ii)-$matrix element of the diagonal $\hat{\beta}_{1}$ matrix and we can write in the general case that $[\beta_{n+1}]_{ij} = \delta_{ij} 
[\beta_{n+1}]_{ii} =\delta_{ij} [\beta_{n+1}]_{i}$ and $[\beta_{1}]_{ij} = \delta_{ij} [\beta_{1}]_{ii} = \delta_{ij} [\beta_{1}]_{i}$. This leads to the following 
analytical expression for the $(ij)-$matrix elements of the $\hat{a}_{n+1}$ matrix
\begin{eqnarray} 
 & & \Bigl[ \hat{a}_{n+1} \Bigr]_{ij} = - 2 \sum_{k} \frac{W_{ik}}{[\beta_{1}]_{i} + [\beta_{1}]_{k} + 2 n} \cdot \frac{W_{kj}}{[\beta_1]_{k} + [\beta_1]_{j} + 2 n} 
   \; \; \; \label{eqzzm1} \\
 &=& - 2 \sum_{k} \frac{1}{K_{i} + K_{k} + 2 n + 3 N_e - 1} \Bigl[ W_{ik} W_{kj} \Bigr] \frac{1}{K_{k} + K_{j} + 2 n + 3 N_e - 1} \nonumber
\end{eqnarray}
where $K_{i}(= K_{ii})$ are the matrix elements of the diagonal $\hat{K}$-matrix (the matrix of hypermomentum) and $n$ is the hyper-radial quantum number which is always 
integer and non-negative. Sometimes the quantum number $n$ is called the number (or index) of hyper-radial excitations (or, simply - the excitation number). Formally, 
this formula is a direct generalization of the Bohr's formula for the energies of bound states in one-electron hydrogen atoms/ions to atoms/ions which contain $N_e$ bound 
electrons. For $N_e = 1$ the formula Eq.(\ref{eqzzm1}) exactly coincides with the Bohr's formula (in atomic units). Indeed, in this case $3 N_e - 1 = 2$, $\hat{W}_{ij} 
= -Q \delta_{ij}, K_i = K_j = \ell$ and $\ell$ is the good quantum number. Therefore, from Eq.(\ref{eqzzm1}) one finds $E_i = \Bigl[ \hat{a}_{n+1} \Bigr]_{ii} = - 
\frac{Q^2}{2 (\ell + 1 + n)^2} = - \frac{Q^2}{2 (n_p)^{2}}$, i.e. the well-known formula for the bound energy levels of one-electron atom/ion (here $n_p = \ell + 1 + n$ 
is the principal quantum number. 
 
Note that for one-electron atoms/ions the energy spectrum of bound states is determined from Eq.(\ref{eqzzm1}) without any reference to the wave functions. However, to 
obtain the bound state spectra in few- and many-electron atomic systems we need to determine the wave functions as well \cite{Fro1}, otherwise the procedure cannot be 
closed. To construct the bound state wave functions for a given atom/ion, which contains $N_e$ bound electrons, we chose a certain atomic term which is described by the 
corresponding quantum numbers, e.g., $\Bigl[ L, M, S, S_z, \pi \Bigr]$ \cite{Fro1}. For this atomic term we construct the system of `physical' HH (hyperangular basis). It 
is assumed that all `physical' HH have the correct permutation symmetry. In the basis of physical HH we calculate all elements of the matrix of the Coulomb potential, i.e. 
the matrix $\hat{W}$. By using the $\hat{W}$ matrix we determine the matrix elements of the $n$-following matrices $\hat{{\cal A}}(n)$ \cite{Fro1}, where $n$ = 0, 1, 2, 
$\ldots$:
\begin{eqnarray}
  [\hat{{\cal A}}(n)]_{ij} &=& \frac{\Gamma(K_i + K_j + 2 n + 3 N_e)}{\sqrt{\Gamma(2 K_i + 2 n + 3 N_e) \Gamma(2 K_j + 2 n + 3 N_e)}} \cdot 
  \frac{2 W_{ij}}{K_{i} + K_{j} + 3 N_e - 1 + 2 n} \; \; \label{wf158a} \\
 &=& \frac{(K_i + K_j + 2 n + 3 N_e - 1)!}{\sqrt{(2 K_i + 2 n + 3 N_e - 1)! (2 K_j + 2 n + 3 N_e - 1)!}} \cdot \frac{2 W_{ij}}{K_{i} + K_{j} + 3 N_e - 1 + 2 n} \nonumber  
\end{eqnarray}
where $n = 0, 1, 2, \ldots$, All these matrices $[\hat{{\cal A}}(n)]$ are symmetric and all their eigenvalues are negative. To obtain the total energies of the bound 
states in one atomic term (e.g., the $\Bigl[ L, M, S, S_z, \pi \Bigr]$ atomic term) we need determine the lowest eigenvalue $\lambda_{1}(n)$ of each of the matrices 
$\hat{{\cal A}}(n)$ for $n = 0, 1, 2, \ldots$). The total energies $E(n)$ of the corresponding bound states in the atom/ion with $N_e$ bound electrons are simply related 
to the $\lambda_1(n)$ eigenvalues by the formula $E(n) = -\frac12 \lambda^{2}_1(n)$ for $n = 0, 1, 2, \ldots$. Note that in this procedure we need only the lowest 
eigenvalue of each of the $\hat{{\cal A}}(n)$ matrices. The set of computed $E(n)$ values (for $n = 0, 1, 2, \ldots$) is the energy spectrum of bound atomic states which 
belong to the same atomic term $\Bigl[ L, M, S, S_z, \pi \Bigr]$. We can designate these energies as the $E_{L, M, S, S_z, \pi}(n)$ values ($n = 0, 1, 2, \ldots$). As 
soon as we know the energies of all bound states in one atomic term, then we have to repeat our procedure for other atomic terms possible for the same $N_e-$electron 
atom/ion. Finally, we obtain the complete bound state spectrum of the $N_e$-electron atom/ion. The procedure for obtaining the corresponding hyper-radial part of the total 
wave functions is described in \cite{Fro1}. 

An obvious advantage of our current approach follows from the fact that all bound state energies are determined in a closed analytical form. These energies are determined 
as the solutions of a number of eigenvalue problems which are closely related to each other. In fact, the corresponding matrices are related to each other by a very 
simple substitution $n \rightarrow n + 1$ in the same formula Eq.(\ref{wf158a}) for all matrix elements. This allows one to investigate the explicit dependencies of the 
total energies of bound states, which belong to the same atomic term, upon the conserving quantum numbers, e.g., upon the number $n$ of excitations (number of radial, or
hyper-radial excitations). Also, we do not need to solve any hyper-radial, or combined (hyper-radial + hyperangular) eigenvalue problem. In other words, the hyper-radial 
dependence of the actual wave function of the $N_e-$electron atom/ion is uniformly determined by the corresponding hyperangular matrix of the potential energy $\hat{W}$, 
numbers of hypermomentum (the numbers $K_i$ and $K_j$ in Eq.(\ref{wf158a})) and by a conserving quantum number of hyper-radial excitations $n$. In many studies, the 
quantum number of excitations (or number of radial/hyper-radial excitations) is designated by $n_r$ (see, e.g., the next Section). Also, in our method we always deal with 
the first-order differential equations. There are other advantages of our procedure based on the exact matrix factorization, but here we do not want to discuss them. 

\section{The hyper-radial O(2,1)-algebra and its representations}

In this Section we discuss another approach which can be considered as some alternative to the method of matric factorization developed above (see also \cite{Fro1}) and 
can essentially be used for the same purposes. The new method is based on the existence of hyper-radial O(2,1)-algebra which is formed by the three self-conjugate operators 
$S, T$ and $U$ defined below. This algebra is resonsible for explicit separation of the discrete and continuous parts of energy spectrum in arbitrary few- and many-body 
Coulomb systems, e.g., in atoms, ions and molecules. Let us define the three following hyper-radial operators 
\begin{eqnarray}
  S = \frac12 r \Bigl( p_{r}^2 + \frac{ K^{2}_{N_e}(\Omega)}{r^2} + 1 \Bigr) \; \; \; , \; \; \; T = r p_r \; \; \; , \; \; \; 
  U = \frac12 r \Bigl( p_{r}^2 + \frac{ K^{2}_{N_e}(\Omega)}{r^2} - 1 \Bigr) \; \; \; \label{AFrol} 
\end{eqnarray}
where $p_{r} = - \imath \frac{\partial}{\partial r}$ and the hyper-radial operator $K^{2}_{N_e}(\Omega)$ is defined above in Eq.(\ref{HH1}).  

Following \cite{FroFirst} let us show that these three operators form the non-compact O(2,1)-algebra, i.e. they obey the following commutation relations
\begin{eqnarray}
  [ S, T ] = - \imath U \; \; \; , \; \; \; [ T, U ] = \imath S \; \; \; , \; \; \; [ U, S ] = - \imath T \; \; \; \label{AFrol1} 
\end{eqnarray}
Furthermore, the Casimir operator of the second order $C_2$ for this non-compact three operator algebra equals 
\begin{equation}
  C_2 = S^{2} - T^{2}- U^{2} = K^{2}_{N_e}(\Omega) = \Lambda^2_{N_e}(\Omega) + \Bigl(\frac{3 N_e - 1}{2}\Bigr)^2 - \Bigl(\frac{3 N_e - 1}{2}\Bigr) \label{Kasimir}
\end{equation}
where all notations used in Eqs.(\ref{AFrol}) - (\ref{Kasimir}) are exactly the same as in Eq.(\ref{HHH}). 

The proof of this theorem is straighforward and based on direct calculations of all three commutators in Eqs.(\ref{AFrol1}) and calculation of the Casimir operator  
$C_2$, Eq.(\ref{Kasimir}). In particular, for the $[ S, U ]$ commutator we can write
\begin{equation}
 [ S, U ] = \frac14 \Bigl\{ [r, r p^{2}_r ] - [ r p^{2}_r, r ] \Bigr\} = \frac12 r [r, r p^{2}_r ] = \imath r p_r = \imath T \; \; \; \label{com1}  
\end{equation}
where $p_r = (-\imath) \frac{\partial}{\partial r}$. For the $[ T, U ]$ commutator one finds
\begin{equation}
 [ T, U ] = \frac12 [r p_r, r p^{2}_r ] + \frac12 [ r p_r, \frac{K^2_{N_e}(\Omega)}{r} ] - \frac12 [ r p_r, r ] = \imath \frac12 r p^{2}_r + 
 (-\imath) (-1) r \frac{K^2_{N_e}(\Omega)}{r^{2}} + \imath r = \imath S \; \; \; \label{com2}  
\end{equation}
where $K^2_{N_e}(\Omega) = \Lambda^2_{N_e}(\Omega) + \Bigl(\frac{3 N_e - 1}{2}\Bigr)^2 - \frac{3 N_e - 1}{2}$. Analogously, for the $[ T, S ]$ commutator we have
\begin{equation}
 [ T, S ] = \frac12 [r p_r, r p^{2}_r ] + \frac12 [ r p_r, \frac{K^2_{N_e}(\Omega)}{r} ] + \frac12 [ r p_r, r ] = \imath \frac12 r p^{2}_r + 
 (-\imath) (-1) r \frac{K^2_{N_e}(\Omega)}{r^{2}} - \imath r = \imath U \; \; \; \label{com3}  
\end{equation}
The Casimir operator $C_2$ is 
\begin{eqnarray}
   C_2 &=& S^{2} - T^{2}- U^{2} = - T^{2}- U^{2} + S^{2} = - r p_r r p_r - \frac14 r \Bigl( p_{r}^2 + \frac{ K^{2}_{N_e}(\Omega)}{r^2} - 1 \Bigr) r (-1) \nonumber \\
 &+& \frac14 r \Bigl( p_{r}^2 + \frac{ K^{2}_{N_e}(\Omega)}{r^2} + 1 \Bigr) r - \frac14 (- r) \Bigl( p_{r}^2 + \frac{ K^{2}_{N_e}(\Omega)}{r^2} - 1 \Bigr) r
   \frac14 r \Bigl( p_{r}^2 + \frac{ K^{2}_{N_e}(\Omega)}{r^2} + 1 \Bigr) r  \nonumber \\
 &=& K^{2}_{N_e}(\Omega) - r p_r r p_r + \frac12 r p_{r}^2 r + \frac12 r^{2} p^{2}_r = K^{2}_{N_e}(\Omega) \; \; \; \label{Cas}
\end{eqnarray}
since
\begin{equation}
  - r p_r r p_r + \frac12 r p_{r}^2 r + \frac12 r^{2} p^{2}_r = \frac12 r^{2} p_{r}^2 + (-\imath) r p_r + \frac12 r^{2} p_{r}^2 - r^{2} p^{2}_r  + \imath r p_r = 0
\end{equation}
The proof of this theorem is over. Note that the exact coincidence of the Casimir operator $C_2$ of this hyper-radial, non-compact O(2,1)-algebra with the analogous 
Casimir operator of the compact O(3 $N_e)$-algebra of hyperangular rotations in the $3 N_e-$dimensional space is not a random fact \cite{Mosh}, but here we cannot 
discuss such a coincidence of the two different Casimir operators and numerous consequencies of this fact. 

The same theorem about O(2,1)-algebra can be proved in the case when the three operators $S, T$ and $U$ are written in their mixed 
matrix-operator forms (in $K-$representation). Indeed, by following our approach developed above one can show that the three following matrix-operators
\begin{eqnarray}
  S &=& \frac12 r \Bigl[ p_{r}^2 + \frac{\Bigl(\hat{K} + \frac{3 N_e + 1}{2}\Bigr) \Bigl(\hat{K} + \frac{3 N_e - 1}{2} - 1 \Bigl)}{r^2} + 1 \Bigr] \; \; \; , \; 
  T = r p_r \; , \nonumber \\
 U &=& \frac12 r \Bigl[ p_{r}^2 + \frac{\Bigl(\hat{K} + \frac{3 N_e + 1}{2}\Bigr) \Bigl(\hat{K} + \frac{3 N_e - 1}{2} - 1 \Bigl)}{r^2} - 1 \Bigr] \; \label{AFrolM} 
\end{eqnarray}
form the non-compact $O(2,1)-$algebra, i.e. they obey the commutation relations, Eq.(\ref{AFrol1}), and the Casimir operator of the second order $C_2$ for this algebra
equals 
\begin{eqnarray}
   C_2 = S^{2} - T^{2} - U^{2} = \Bigl(\hat{K} + \frac{3 N_e + 1}{2}\Bigr) \Bigl(\hat{K} + \frac{3 N_e - 1}{2} - 1 \Bigl)  \; \; \label{CasM}
\end{eqnarray}

This theorem allows us to perform a number of advanced transformations of the Coulomb Hamiltonian(s) (one example is discussed below) and explicitly construct different 
series of representations of the non-compact O(2,1)-algebra. Briefly, it is possible to say that this theorem provides a very powerful tool for investigation of the bound 
state spectra of different Coulomb systems. In particular, based on this theorem one can separate the bound and continuous (or unbound) parts of the `united' energy 
spectrum in an arbirary Coulomb system, including few- and many-electron atoms, ions and molecules. To achieve this goal let us multiply the original Schr\"{o}dinger 
equation $(H(r, \Omega) - E) \Phi(r, \Omega) = 0$ by the hyper-radius $r$ from its left, i.e. 
\begin{eqnarray} 
  0 = r (H(r, \Omega) - E) \Phi = \Bigl\{\Bigl[ \Bigl(\frac{1}{2 m_e} - E\Bigr) S + \Bigl(\frac{1}{2 m_e} + E\Bigr) U\Bigr] + W(\Omega) \Bigr\} \Phi(r, \Omega) 
 \label{Schr0}
\end{eqnarray} 
where $W(\Omega)$ is the angular part of the potential energy and $m_e$ is the electron mass, or, in general, the mass of an arbitrary electrically charged particle bound 
(by a Coulomb potential) to an infinitely heavy center. Now, we can introduce the new function $\Psi(r, \Omega) = \exp(\imath \chi T) \Phi(r, \Omega)$, where $\chi$ is
somenumerical (real) parameter (angle), while the operator $T$ is defined in Eq.(\ref{AFrol1}) \cite{Virial}. Since the angle $\chi$ is real, then the transformation 
$\exp(\imath \chi T)$ is unitary and the functions $\Psi(r, \Omega)$ and $\Phi(r, \Omega)$ are related to each other by a unitary transformation. In the basis of new 
functions $\Psi(r, \Omega)$ one finds for the operator $r (H(r, \Omega) - E)$ from Eq.(\ref{Schr0}):
\begin{eqnarray} 
 \exp(-\imath \chi T) r (H(r, \Omega) &-& E) \exp(\imath \chi T) = [ \Bigl(\frac{1}{2 m_e} - E\Bigr) \Bigl( S \cosh\chi + U \sinh\chi ) + W(\Omega) \nonumber \\
 &+& \Bigl(\frac{1}{2 m_e} + E\Bigr) \Bigl( U \cosh\chi + S \sinh\chi ) \bigr] \label{Schr1}
\end{eqnarray}  
The formula for the expression in the right-hand side of Eq.(\ref{Schr1}) is reduced to the form 
\begin{eqnarray}
    \Bigl(\frac{1}{2 m_e} - E\Bigr) \Bigl( S \tanh\chi + U ) + \Bigl(\frac{1}{2 m_e} + E\Bigr) \Bigl( U + S \tanh\chi ) + \frac{W(\Omega)}{\cosh\chi} \label{Schr2}
\end{eqnarray}
Therefore, if we chose $\tanh\chi$ in the form 
\begin{equation}
  \tanh\chi = \frac{\frac{1}{2 m_e} + E}{\frac{1}{2 m_e} - E} \label{choice1}
\end{equation}
then we can exclude the operator $U$ (or generator $U$) from Eq.(\ref{Schr2}). Note that for real angles $\chi$ the numerical value of $\tanh\chi$ is always less than unity. 
Therefore, our choice of $\tanh\chi$ in Eq.(\ref{choice1}) corresponds to the bound (or discrete) spectrum in Coulomb systems for which $E < 0$. Analysis of other cases and 
interaction potentials can be found in \cite{FroFirst}. 

Now, by using the relation $\frac{1}{\cosh\chi} = \sqrt{1 - \tanh^{2}\chi}$ (see, e.g., \cite{GR}), Eq.(\ref{Schr2}) and Eq.(\ref{choice1}) we reduce the Schr\"{o}dinger 
equation, Eq.(\ref{Schr0}), to the following final form
\begin{equation}
  \Bigl[ \sqrt{- \frac{2 E}{m_e}} \; \; S + W(\Omega) \Bigr] \Psi(r, \Omega) = 0 \label{Schrof}
\end{equation}
which has the discrete spectrum only. In the basis of hyperspherical harmonics (HH) this equation has the following (equivalent) matrix-operator form 
\begin{equation}
  \Bigl[ \sqrt{- \frac{2 E}{m_e}} \; \; \hat{S}(r) + \hat{W} \Bigr] \vec{\Psi}(r) = 0 \label{Schrofa}
\end{equation}
where the matrix $\hat{S}(r)$ is the differential operator of the second-order in respect to the hyper-radius $r$. In contrast with this, the potential matrix $\hat{W}$ 
does not depend upon $r$. The theorem about the discrete (positive) series $D^{+}$ of representations of the O(2,1)-algebra (see, e.g., \cite{Barut}, \cite{Barut1} and 
\cite{Perelom}) states that: there is a normalized basis of $\mid n K \rangle$-states such that 
\begin{eqnarray}
 S \mid n K \rangle = n \mid n K \rangle \; \; \;  \label{theor1}
\end{eqnarray}
and 
\begin{eqnarray}
 C_2 \mid n K \rangle = K^{2}_{N_e}(\Omega) \mid n K \rangle = \Bigl(\hat{K} + \frac{3 N_e + 1}{2}\Bigr) \Bigl(\hat{K} + \frac{3 N_e - 1}{2} - 1 \Bigl)
 \mid n K \rangle \; \; , \; \label{theor2}
\end{eqnarray}
where $n = K + \frac{3 N_e - 1}{2}, K + \frac{3 N_e - 1}{2} + 1, K + \frac{3 N_e - 1}{2} + 2, \ldots$, or $n = K + \frac{3 N_e - 1}{2} + n_r$, and $n_r = 0, 1, 2, \ldots$. 
In general, these $\mid K n \rangle$ states form the basis for the $D^{+}$ series of representations of the non-compact O(2,1)-algebra. Furthermore, for one-electron hydrogen 
atom and hydrogen-like ions, when $W(\Omega) = Q$ (in atomic units $e^{2} = 1$) and $K = \ell$ is an observable (or conserving) quantum number, these $\mid n K \rangle$-states 
(or $\mid n \ell \rangle$-states) exactly coincide with the radial parts of the actual wave functions. With these numerical values one finds from the formula Eq.(\ref{Schrof}) 
that $E = - \frac{m_e Q^2}{2 n^2}$, where $n = \ell + 1 + n_r$ is the principal quantum number (conserved), $\ell$ is the quantum number of angular momentum (conserved) and 
$n_r$ is the hyper-radial quantum number (or number of hyper-radial excitations) which is also conserved in one-electron atomic systems. Thus, based on the Schr\"{o}dinger 
equation Eq.(\ref{Schrof}) we have re-derived the Bohr's formula for the energy levels of an arbitrary one-electron atomic system. The radial parts of the bound state wave 
funcions in such systems excatly coincide with the corresponding unit-norm $\mid n_r \ell \rangle$-states, or with the identical $\mid n \ell \rangle$-states, where $n = n_r 
+ \ell + 1$, mentioned above.   
 
For few- and many-electron atomic systems with $N_e \ge 2$ the values of hypermomentum $K_{i}$ are not conserving quantum numbers and, therefore, we need to develop a different 
technique. First, we need to derive the explicit formula for the spectrum of operator $S$ and corresponding eigenfunctions. This will lead us to the explicit formula for the 
hyper-radial $\mid n K \rangle-$states. These unit-norm hyper-radial $\mid n K \rangle-$states form the basis for the $D^{+}$-series of representations of the of hyper-radial 
O(2,1)-algebra. The operator $S$ is included in the Schr\"{o}dinger equation, Eq.(\ref{Schrof}). It is clear that the scalar (or one-dimensional) radial equation $S \phi(r) = n 
\phi(r)$ (or $(S - n) \phi(r) = 0$), which was used for one-electron atomic systems, must be replaced by an infinite-dimensional matrix equation $(\hat{S}(K) - \hat{n}) 
\vec{\Phi}(r) = 0$, where $\hat{S}(K)$ is the diagonal matrix of the $S$ operator in $K-$representation, while the notation $\hat{n}$ stands for infinite-dimensional matrix 
$\hat{n}(\hat{K})$ of the principal quantum numbers and $\vec{\Phi}(r)$ is the corresponding eigenvector which is the function of hyper-radius $r$ only (each element of this 
vector depends upon the hyper-radius $r$ only). In general, each matrix element of the $\hat{n}$ matrix must be a uniform function of the corresponding hyperspherical momentum 
$K$, i.e. each $n_{i}$ value depends upon the unique $K_i$ value (for each $i$) and vice versa. As follows from here the matrix $\hat{n}$ is also a diagonal matrix in 
$K-$representation. For the diagonal matrices $\hat{n}$ and $\hat{K}$ we always have $n_{ii} = n_{ii}(K_{ii})$ (or $n_{i} = n_{i}(K_{i})$), where $i$ = 1, 2, 3, $\ldots$. The 
explicit form of the matrix equation $(\hat{S}(K) - \hat{n}) \vec{\Phi}(r) = 0$ is 
\begin{equation}
   \Bigl[ -\frac{\partial^{2}}{\partial r^{2}} + \frac{\hat{K}(\hat{K} + 1)}{2} + 1  - \frac{2 \hat{n}}{r} \Bigr] \vec{\Phi}(r) = 0 \; \; \label{meq1}
\end{equation} 
where $\hat{K}$ and $\hat{n}$ are the diagonal matrices, the vector $\vec{\Phi}(r)$ is regular at $r = 0$ and actual dimension of these matrices coincides with the number of 
hyperspherical harmonics (HH) used.  
 
To solve Eq.(\ref{meq1}) we represent the infinite-dimensional eigenvector $\vec{\Phi}(r)$ from Eq.(\ref{meq1}) in the form 
\begin{equation}
    \vec{\Phi}(r) = r^{\hat{\gamma}} \vec{\Psi}(r) \exp(-\alpha r) \; \; \label{meq2}
\end{equation}
where $\hat{\gamma}$ is the matrix which is diagonal in $K-$representation, while $\alpha$ is a positive numerical constant. i.e., it is a scalar, or $c-$number. Substitution 
of the wave function $\vec{\Phi}(r)$ written in the form of Eq.(\ref{meq2}) into Eq.(\ref{meq1}) after a few simple transformations leads to the following equation
\begin{equation}
   r \frac{d^2 \vec{\Psi}(y)}{d r^2} - (2 \hat{K} + 3 N_e - 1 - r) \frac{d \vec{\Psi}(y)}{d r} - (\hat{K} + \frac{3 N_e - 1}{2} - \hat{n}) \vec{\Psi}(y) = 0 \; \; \label{meq3}
\end{equation} 
where $y = 2 r$. By performing these transformations we also found that $\hat{\gamma} = \hat{K} + \frac{3 N_e - 1}{2}$ and $\alpha = 1$. The last equation, Eq.(\ref{meq3}), 
coincides with the corresponding equation for the confluent hypergeometric functions ${}_{1}F_{1}(a, b; z)$ which is explicitly defined in \cite{GR}. However, the coefficients 
of this equation are the matrices (infinite-dimensional matrices!) $\hat{a} = \hat{K} + \frac{3 N_e - 1}{2} - \hat{n}$ and $\hat{b} = 2 \hat{K} + 3 N_e - 1$. In reality, it is 
difficult to operate with the confluent hypergeometric functions of matrix arguments. However, as mentioned above the both matrices $\hat{K}$ and $\hat{n}$ in Eq.(\ref{meq3}) 
are diagonal. Therefore, we can re-write the same equation in the following one-dimensional (or scalar) form for each component of the $\vec{\Psi}(y)$ vector-function:
\begin{equation}
   r \frac{d^2 \Psi_{i}(y)}{d r^2} - (2 K_{i} + 3 N_e - 1 - r) \frac{d \Psi_{i}(y)}{d r} - (K_{i} + \frac{3 N_e - 1}{2} - n_{i}) \Psi_{i}(y) = 0 \; \; \label{meq35}
\end{equation} 
where $i$ = 1, 2, 3, $\ldots$, $y = 2 r$, while $K_{i} = K_{ii}$ and $n_{i} = n_{ii}$ are the corresponding diagonal matrix elements of the two matrices $\hat{K}$ and 
$\hat{n}$, respectively. Solution of this equation, Eq.(\ref{meq35}), is the confluent hypergeometric function: ${}_{1}F_{1}(K_i + \frac{3 N_e - 1}{2} - n_i, 2 K_i +  3 N_e - 
1; 2 r)$. The finite-norm solutions of this equation, Eq.(\ref{meq35}), do exist, if (and only if) the following condition for the matrix elements is obeyed: $\hat{K}_{ii} + 
\frac{3 N_e - 1}{2} - \hat{n}_{ii} = - n_r$, or $K_{i} + \frac{3 N_e - 1}{2} - n_{i} = - n_r$, where $n_r$ is also a non-negative integer number which does not depend upon the 
index $i$. The second coefficient ($b$) of the confluent hypergeometric functions ${}_{1}F_{1}(a, b; y)$ function equals $2 K_i +  3 N_e - 1$, i.e. it is an integer, positive 
number. Therefore, the finite-norm solution of Eq.(\ref{meq35}) coincides with the generalized Laguerre polynomials (see, e.g., \cite{GR}, \cite{AS}) $L^{k_i}_{n_r}(2 r)$, 
where $k_i = 2 K_i +  3 N_e - 1$ and $n_r = n_i - K_{i} - \frac{3 N_e - 1}{2}$. Now, for the $i-$th component of the $\vec{\Phi}(r)$ vector one finds
\begin{equation}
 \Phi_{i}(r) = r^{K_{i} + \frac{3 N_e - 1}{2}} \cdot L^{2 K_i +  3 N_e - 1}_{n_r}\Bigl( 2 r \Bigr) \cdot \exp(- r) \label{meq05}
\end{equation} 
The set of hyper-radial functions, defined by Eq.(\ref{meq05}) is complete in the ${\cal L}^{2}_{r}(0, +\infty)$ Hilbert space. 

By using the explicit form of the obtained hyper-radial $\Phi_{i}(r)$ function it is relatively easy to check that each of these functions (for each $i$) obeys the two following 
conditions: $S \Phi_{i}(r) = n_{i} \Phi_{i} = (n_r + K_{i} + \frac{3 N_e - 1}{2}\Bigl) \Phi_{i}(r)$ and $C_2 \Phi_{i}(r) = (S^{2} - T^{2} - U^{2}) \Phi_{i}(r) = \Bigl(\hat{K} 
+ \frac{3 N_e + 1}{2}\Bigr) \Bigl(\hat{K} + \frac{3 N_e - 1}{2} - 1 \Bigl) \Phi_{i}(r)$. In other words, the hyper-radial functions $\Phi_{i}(r)$ are proportional to the 
corresponding $\mid n_i, K_i \rangle$-state (or $\mid n_{r}, K_i \rangle$-states) mentioned above. These $\mid n_{r}, K_i \rangle$-states (or $\mid n_i, K_i \rangle$-states, 
where $n_i = K_i + \frac{3 N_e - 1}{2} + n_r$) form the basis of hyper-radial, unit-norm functions for the discrete and positive $D^{+}$-series of representation of the non-compact 
O(2,1)-algebra. The exact coincidence of the $\mid n_{r}, K_i \rangle$ states with the hyper-radial $\Phi_{i}(r)$ functions is observed, if (and only if) such functions have unit 
norm, i.e. 
\begin{equation}
 \mid n_{r}, K_i \rangle = \Phi_{i}(r) = B^{n_r}_{K_i}(N_e) \cdot r^{K_{i} + \frac{3 N_e - 1}{2}} \cdot L^{2 K_i +  3 N_e - 1}_{n_r}\Bigl( 2 r \Bigr) 
 \cdot \exp(- r) \; \; \; \label{meq055}
\end{equation} 
where the notation $B^{n_r}_{K_i}(N_e)$ stands for the corresponding normalization factor. The explicit formula for this normalization factor is  
\begin{equation} 
   B^{n_r}_{K_i}(N_e) = 2^{-K_i - \frac{3 N_e}{2}} \sqrt{\frac{n_r !}{(2 K_i + 3 N_e - 1 + n_r)!}}  \; \; \;  \label{norm}
\end{equation} 
Now, the eigenvector, Eq.(\ref{meq1}), is written in the following form (in matrix notation)
\begin{eqnarray}
  \vec{\Phi}_{n_{r}}(r) &=& {\vec C} \cdot  B^{n_r}_{\hat{K}}(N_e) \cdot r^{\hat{K} + \frac{3 N - 1}{2}} \cdot {}_{1}F_{1}(- n_{r}, 2 \hat{K} + 3 N_e - 1; 2 r \Bigr) \exp(- r) 
  \nonumber \\
   &=& r^{\hat{K} + \frac{3 N - 1}{2}} \cdot L^{2 \hat{K} + 3 N_e - 1}_{n_r}\Bigl( 2 r \Bigr) \exp(- r) \; \; \label{meq25}
\end{eqnarray}
where ${\vec C} = (C_1, C_2, C_3, \ldots$) is the vector of unknown linear coefficients which concide with the corresponding expansion coefficients of the actual state in the 
$N_e$-electron atom/ion represented as a linear combination of the basis $\mid n_{r}, K_i \rangle$-states. In mathematical physics such linear coefficients are often called the 
generalized Fourier coefficients. As mentioned above these unit-norm states form the basis for the $D^{+}$ series of representations of the non-compact O(2,1)-algebra. The fact 
that to approximate any actual bound state wave function in an atom/ion with $N_e-$bound electrons one needs to use only the basis of $\mid n_{r}, K_i \rangle$-states indicates 
clearly the leading role of the `hidden' O(2,1)-symmetry which can be found in an arbitrary many-particle Coulomb system. This also shows that there is no principal difference 
between the bound state spectra in one-electron hydrogen-like atomic systems and in arbitrary many-electron atoms and ions. In contrast with this, for few- and many-particle 
nuclear systems, where particles (or nucleons) interact with each other by non-Coulomb forces, such a simple relation between the bound state spectra of different systems does 
not exist.

Thus, the original Schr\"{o}dinger equation, Eq.(\ref{Schrof}), is reduced to the following `operator' form
\begin{equation}
  \Bigl[ \sqrt{- \frac{2 E}{m_e}} + S^{-\frac12} W(\Omega) S^{-\frac12} \Bigr] \Psi(r, \Omega) = 0 \; \; \; , \; \label{Schrof1}
\end{equation}
or, in the equivalent matrix-operator form, Eq.(\ref{Schrofa}): 
\begin{equation}
  \Bigl[ \sqrt{- \frac{2 E}{m_e}} + \hat{S}^{-\frac12} \hat{W} \hat{S}^{-\frac12} \Bigr] \mid \Psi \rangle = 0 \; \; \; , \; \label{SchrofAa}
\end{equation}
where the unknown wave function $\mid \Psi \rangle$ ($ket-$vector \cite{Dir}) is represented as a linear combination of the $\mid n_{r}, K_i \rangle$-states. In the basis of the 
unit-norm $\mid n_{r}, K_i \rangle$-states one can reduce this Schr\"{o}dinger equation, Eq.(\ref{SchrofAa}), to the following eigenvalue problem which is written below in the 
matrix form 
\begin{equation}
 \Bigl[ \sqrt{- \frac{2 E}{m_e}} \delta_{ij} + C_i \frac{B^{n_r}_{K_i}(N_e) \cdot W_{ij} \cdot B^{n_r}_{K_j}(N_e)}{\sqrt{(K_{i} + \frac{3 N_e - 1}{2} + n_r) (K_{j} + 
 \frac{3 N_e - 1}{2} + n_r)}} C_j \Bigr] = 0 \; \; \; , \; \label{SchrofA}
\end{equation}
Solution of this equation is equivalent to the following variational problem
\begin{equation}
   \sqrt{- \frac{2 E}{m_e}} = - \max_{{\vec C}} \langle {\vec C} \mid \hat{M} \mid {\vec C} \rangle \label{variat}
\end{equation}
where ${\vec C}$ are the unit-norm, numerical (real) vectors and the matrix $\hat{M}$ has the following matrix elements 
\begin{equation}
 M_{ij}(n_r) = \frac{B^{n_r}_{K_i}(N_e) \cdot W_{ij} \cdot B^{n_r}_{K_j}(N_e)}{\sqrt{\Bigl(K_{i} + \frac{3 N_e - 1}{2} + n_r \Bigr) (K_{j} + \frac{3 N_e - 1}{2} + n_r \Bigr)}} 
 \; \; \; \label{Mmatrix}
\end{equation}
Solution of this eigenvalue problem allows one to obtain the corresponding $n_r-$th energy level and determine the linear coefficients ${\vec C}$ which are needed to construct 
the total wave functions $\Psi(r, \Omega)$ by representing them as linear combinations of the products of `physical' hyper-spherical harmonics and $\mid n_r, K_{i} + 
\frac{3 N_e - 1}{2} \rangle$-states. If $\lambda_{n_{r}}$ is the minimum/maximum of the right-hand side of Eq.(\ref{variat}), then we can write the following formula for the 
total energy $E_{n_{r}}$ of the corresponding atomic level $E_{n_{r}} = -\frac12 \lambda^{2}_{n_{r}}$, where $m_e = 1$ in atomic units and $n_r$ is the `index of excitation' 
of this level ($n_r = 0, 1, 2, \ldots$). By taking into accout the explicit form of the matrix elements ($M_{ij}$) of the $\hat{M}$ matrix we also find another formula for the 
$E_{n_{r}}$ value
\begin{eqnarray}
 E_{n_{r}} = -\frac12 \max_{{\vec C}} \langle {\vec C} \mid  \frac{B^{n_r}_{K_i}(N_e) \cdot W_{ik} \cdot \Bigl(B^{n_r}_{K_k}(N_e)\Bigr)^{2} \cdot W_{kj} \cdot 
 B^{n_r}_{K_j}(N_e)}{\sqrt{ K_{i} + \frac{3 N_e - 1}{2} + n_r} \Bigl( K_{k} + \frac{3 N_e - 1}{2} + n_r \Bigr) \sqrt{ K_{j} + \frac{3 N_e - 1}{2} + n_r}} \mid {\vec C} 
 \rangle \; \; \; \label{MMmatrix}
\end{eqnarray}
where each of the trial vectors ${\vec C}$ has the unit norm.

\section{Discussion and Applications}

Note that in the both approaches described in this study the total energies of the corresponding bound states $E_{n_{r}}$ are obtained from the relations $E_{n_{r}} = -\frac12 
\lambda^{2}_{n_{r}}$, where $\lambda^{2}_{n_{r}}$ is the maximal (by the absolute value) eigenvalue of some matrix. For the matrix factorization method (see Section II) 
$\lambda_{n_{r}}$ is such an eigenvalue of the $\hat{{\cal A}}(n_r)$ matrix defined by Eq.(\ref{wf158a}), while in the method described in this Section $\lambda_{n_{r}}$ is 
the corresponding eigenvalue of the matrix $\hat{M}$ which is defined by Eq.(\ref{Mmatrix}). This directly follows from factorization of the Coulomb Hamiltonian(s) of arbitrary
few- and many-body (or many-electron) systems mentioned above. Furthermore, these two matrices ($\hat{{\cal A}}(n_r)$ and $\hat{M}(n_r)$) are similar to each other as it follows 
from the explicit formulas for their matrix elements. Formally, each of these matrices corresponds to the square root of the actual Coulomb Hamiltonian of an arbitrary atom/ion 
with $N_e$ bound electrons. In other words, for arbitrary Coulomb systems, including actual atoms and ions, we need to determine the maximal (by the absolute value) eigenvalue 
$\lambda_{n_{r}}$ of the matrix which represents the $\sqrt{H}$ operator in the basis of hyperspherical harmonics. It is interesting to note that we do not need to use the 
complete Hamiltonians $H$ to solve the bound state problem for an arbotrary Coulomb few- and many-body system. This remarkable fact directly follows from our result mentioned 
above (see also \cite{Fro1}): the Hamiltonian of an arbitrary Coulomb few- and many-particle system is always factorized. In some sense  this is a physical consequence of the 
well known `additional' symmetry for the Coulomb systems, where all particles interact with each other by the Coulomb potential. Another group of similar systems is formed by 
the `harmonic' few- and many-body systems, where all particles interact with each by the regular harmonic potential $(\simeq A r^{2})$. It can be shown that the Hamiltonian of 
an arbitrary few- and many-body harmonic system is always factorized. Such systems and their bound state spectra will be considered elsewhere.

Unfortunately, for a given (or original) Coulomb Hamiltonian $H$ we cannot determine the operator $\sqrt{H}$ unamibigously. Furthermore, the operator $\sqrt{H}$ and final 
equations for the energy levels of the Coulomb few- and many-body systems are always written in a number of similar (but not identical!) forms. For istance, in this Sections 
we replaced the original Schr\"{o}dinger equation, Eq.(\ref{Schrof}), by the `final' (equivalent) equation Eq.(\ref{Schrof1}). However, it is also possible to consider another 
equivalent equation
\begin{equation}
  \Bigl\{ \sqrt{- \frac{2 E}{m_e}} + \frac12 \Bigl[ S^{-1} W(\Omega) + W(\Omega) S^{-1} \Bigr]\Bigr\} \Psi(r, \Omega) = 0 \; \; \; , \; \label{Schrof1x}
\end{equation}
which can also be used to produce analytical formula(s) for the energy spectra in actual atoms and ions. In general, in Eqs.(\ref{Schrof1}) and (\ref{Schrof1x}) one can also 
applied the symmetrized operator $\frac12 [ S^{-\alpha} W(\Omega) S^{-1 + \alpha} + S^{-1 + \alpha}  W(\Omega) S^{-\alpha} ]$, where $\alpha$ is an arbitrary real number 
bounded between 0 and unity, i.e. $0 \le \alpha \le 1$. All arising equations, Eq.(\ref{Schrof1}), Eq.(\ref{Schrof1x}) and others, are similar to each other, but they do 
not coincide exactly. Nevertheless, their solutions converge to the same answer (or limit) when the dimensions of matrices included in these equations increase to the infinity. 

It should be mentioned here that applications of the both methods developed in this study for accurate numerical computations of a number of few-electron atoms and ions were 
not very successful. The main reson for this is simple: all mehtods based on the hyperspherical harmonics are not appropriate methods for accurate description of the
electron-electron correlations in such systems (for more detalis, see, e.g., \cite{Fro1} and \cite{Fro1986}). In \cite{Fro1} by using the method of matrix factorization 
(see Section II above) the following total energies for some lower-lying bound states in the ${}^{1}S-$term in the He atom with the infinitely heavy nucleus: $E_1$ = -2.9037175 
$a.u.$ (the ground state), $E_2$ = -2.144954 $a.u.$ (the first excited state), $E_3$ = -2.06033 $a.u.$ (the second excited state), $E_4$ = -2.0318 $a.u.$ (the third excited 
state). These results look good, but highly accurate values of these total energies obtained in our earlier calculations for these bound states are \cite{FroHe} (see also 
\cite{Fro2015}): $E_1$ = -2.903724377034119598311159245194405(5) $a.u.$, $E_2$ = -2.145974046054417415(10) $a.u.$, $E_3$ = -2.06127198974090848(5) $a.u.$, $E_4$ = 
-2.03358671703072520(7) $a.u.$ All these energies have been determined with the use of our exponential variational expansion in the perimetric three-body coordinates which 
substantially better represent the actual geometry of the triangle of particles. Analogous calculations of the total bound state energies of the $1^{1}S-$states in the 
two-electron atoms and ions lead to the following results for the ${}^{\infty}$Li$^{+}$ and ${}^{\infty}$Be$^{2+}$ ions: $E_1$(Li$^{+})$ = -7.2799108 $a.u.$ and 
$E_1$(Be$^{2+})$ = -13.655483 $a.u.$ The corresponding highly accurate values of these total energies are: $E_1$(Li$^{+}$) = -7.27991341266930596491875 $a.u.$ and 
$E_1$(Be$^{2+}$) = -13.65556623842358670208051 $a.u.$ \cite{Fro2015} (currently, these results can be improved to much better accuracy). 

These simple examples of two-electron atoms and ions indicate clearly that any variational method based on the use of hyperspherical harmonics cannot be applied for modern, 
highly accurate computations of bound states in few-body systems. However, the both methods developed in this study can be useful for general theoretical analysis of the 
bound state spectra in various few-body systems. This follows from the exact matrix factorization of an arbitrary Coulomb Hamiltonain and analytical separation of the 
discrete and continuous parts of energy spectra in such systems. For instance, let us consider the classification of bound states in the Coulomb few- and many-body systems. 
First, based on the formulas, Eqs.(\ref{Schrof}) - (\ref{meq1}), we can show that the operator $S$ is a compact operator \cite{Rudin}. Indeed, this operatos is a self-adjoint 
operator ($S^{*} = S$ and $S S^{*} = S^{*} S$) and its spectrum has only one limiting point which equals to the ionization (or dissociation) threshold energy (or `physical 
zero'). Furthermore, if $\lambda$ is an arbitrary eigenvalue of $S$, then the corresponding eigenspace is a finite dimensional subspace, i.e. $dim {\cal N}( S - \lambda I) 
< \infty$. Therefore, we can apply theorem (12.30) from \cite{Rudin} which proves that the operator $S$ is a compact operator. 
 
Briefly, this means that classification of the bound state spectra in the Coulomb few- and many-body systems can be performed by applying the classification scheme known (and 
used) for compact operators \cite{Maurin}. To classiify the compact operators one needs to determine the following sums 
\begin{equation}
     S_p(N) = \sum^{N}_{i=1} \mid \lambda_{i} \mid^{p} dim\{\Psi(\lambda_i)\} = \sum^{N}_{i=1} \mid \lambda_{i} \mid^{p} dim\{{\cal H}_{i}\} \; \; \; \label{sums}
\end{equation} 
where $p$ is a non-negative integer number, $\lambda_{i}$ is the corresponding eigenvalue of the Coulomb Hamiltonian $H$ and $dim\{\Psi(\lambda_i)\}$ is the total (algebraic)
dimension of the corresponding eigenspace ${\cal H}_{i}$. The three sums $S_p$ which correspond to the following values of $p$: $p = 0, 1, 2$ are of great interest for the 
classification of the compact operators (and bound state spectra of the Coulomb systems). Briefly, if for some Coulomb Hamiltonian $H$ the sum $S_0(N)$ converge (when 
$N \rightarrow \infty$), then this Hamiltonian has the finite-dimensional spectrum of bound states (or finite discrete spectrum, for short). If the $S_0(N)$ sum diverges at 
$N \rightarrow \infty$, then we are dealing with the infinite-dimensional spectrum of bound state (or infinite discrete spectrum, for short). In general, the Coulomb 
Hamiltonians with the infinite-dimensional spectra of bound states can be separated into the two following classes: (1) the nuclear (or kernel) compact operators for which the 
sums $S_{p=1}$ converge, when $N \rightarrow \infty$, and (2) the Hilbert-Schmidt compact operators for which the sums $S_{p=2}$ converge, but analogous sums $S_{p=1}$ diverge 
when $N \rightarrow \infty$. Examples of the Coulomb few- and many-body systems with the Hilbert-Schmidt spectra of bound states are numerous and include, e.g., all neutral 
atoms and positively charged ions. A large number of the Coulomb few-body and many-electron atomic systems have the finite bound state spectra. In particular, all negatively 
charged ions, e.g., the H$^{-}$ ion and analogous ions of hydrogen isotopes, Li$^{-}$ ion(s), etc, have only one bound state. The finite bound state (or energy) spectra are 
typical among various few-body systems, e.g., the Ps$^{-}$ ion and Ps$_2$ quasi-molecule have one bound state each, while the three-body muonic $p p \mu$ ions has two bound 
states, while the analogous $d t \mu$ and $t t \mu$ ions have five and six bound states, respectively.        

The Coulomb few- and many-body systems which have the kernel (or nuclear) energy spectra are slitgly more difficult to find. However, such systems do exist and one well known
example is the molecular ${}^{\infty}$H$^{+}_2$ ion which has only one bound electron. This one-electron ion has an infinite number of bound states, which are, in general, 
less populaled than the bound state spectra of the helium atom and/or Li-like (three-eletron) ions. This is true for an arbitrary interparticle distance $R$ in the molecular 
${}^{\infty}$H$^{+}_2$ ion. However, when $R = 0$, then we have a `united ion' (i.e. the one-electron helium ion He$^{+}$), and the energy spectrum of bound states becomes the 
Hilbert-Schmidt spectrum. 

To avoid a pure mathematical discussion, let us apply this spectral classification scheme to an arbitrary one-electron atomic system. Consider an atom with the infinitely 
heavy, point nucleus with the positive electric charge $Q e$ and one electron (charge $-e$) bound to this nuclues by the Coulomb potential. The sum $S_{p=1}(N)$ for this 
system is
\begin{equation}
  S_1(N) = \sum^{N}_{i=1} \mid \lambda_{i} \mid dim\{\Psi(\lambda_i)\} = \frac{Q e^{2}}{2} \sum^{N}_{i=1} \Bigl[\frac{1}{n^{2}} \sum^{n-1}_{\ell=0} (2 \ell + 1)\Bigl] = 
 \frac{Q e^{2}}{2} \sum^{N}_{i=1} \frac{n (n - 1) + n}{n^2}   \; \; \; \label{sums1}
\end{equation} 
It is clear that this sum has no finite limit when $N \rightarrow \infty$, i.e. this sum is divergent when $N \rightarrow \infty$. The sum $S_{p=2}(N)$ for the same system 
is
\begin{equation}
  S_2(N) = \sum^{N}_{i=1} \mid \lambda_{i} \mid^{2} dim\{\Psi(\lambda_i)\} = \frac{Q^{2} e^{4}}{4} \sum^{N}_{i=1} \Bigl[\frac{1}{n^{4}} \sum^{n-1}_{\ell=0} (2 \ell + 1)\Bigl] 
 = \frac{Q e^{2}}{2} \sum^{N}_{i=1} \frac{n^2}{n^4} \; \; \; \label{sums2}
\end{equation}   
Finally, the infinite sum $S_2$ is reduced to the following form 
\begin{equation}
  S_2 = \frac{Q^2 e^{4}}{4} \lim_{N \rightarrow \infty} \sum^{N}_{i=1} \Bigl[\frac{1}{n^2} \Bigr] = \frac{Q^2 e^{4}}{4} \cdot \zeta(2) \approx 
   1.6449340668482264 \cdot Q^2 e^{4} \; \; \; \label{sums2A}
\end{equation}  
where $\zeta(n)$ is the Riemann's $\zeta-$function (see, e.g., \cite{GR}) and $\zeta(2) = \frac{\pi^{2}}{6}$ = 1.6449340668482264$\ldots$. Therefore, the limit of the sum $S_2$ 
is finite (at $N \rightarrow \infty$) and we deal with the Hilbert-Schmidt energy spectrum. In general, the energy spectrum of bound states in the hydrogen-like (or one-electron) 
atomic system can be considered as an `almost ideal' example of the Hilbert-Schmidt spectrum. Note also that in all formulas derived in this Section we ignored the known 
experimental fact that the bound states of any hydrogen-like atom/ion are the doublet states. The total electron spin of these doublet states equals $\frac12$, while their spin 
multiplicity equals 2. Therefore, the total sum $S_2$, Eq.(\ref{sums2A}), must be multiplied by the of 2. However, all these `addtional' factors have a restricted (or secondary) 
meaning for our present purposes. The main result is the finite numerical value of the $S_2$ sum. Classification of the bound state spectra of other atomic and molecular systems 
can be performed analogously. 

\section{Conclusion}

In this study we continue our attempts to develop the general theory of bound state spectra in few- and many-body Coulomb systems. In particular, by using the method of 
hyperspherical harmonics and hyper-radial operators written in explicit self-conjugate forms we show that the Hamiltonians of the Coulomb few- and many-particle systems are 
always factorized. Furthermore, such a factorization is exact and unambiguous for an arbitrary Coulomb few- and/or many-body system, including different atoms, ions and molecules. 
These results of our analysis is an important step in our understanding of the bound state spectra of the few- and many-particle Coulomb systems. In particular, in this study all 
hyper-radial operators are written in self-conjugate forms. This allows us to simplify our analysis and make more clear a number of relations between different operators.  

Another interesting direction of research is the use of $D^{+}$-series of representations of the hyper-radial O(2,1)-algebra constructed in this study for the Coulomb systems
with arbitrary number of particles. This approach allows one to obtain the formulas for the corresponding energy levels in each atomic term. The exact wave functions of any 
$N_e-$electron atom/ion is represented as a linear combination of the $\mid n K \rangle-$states from the $D^{+}$-series of representations of the hyper-radial O(2,1)-algebra. 
Furthermore, the actual wave function of an arbitrary $N_e-$electron atom/ion is an optimal variational (linear) combination of the $\mid n K \rangle-$states. No other 
hyper-radial functions are needed to solve this problem.

As follows from the results of our study the method of hyperspherical harmonics has a great potential in applications to the bound state problems in few- and many-electron 
atomic problems. Some of such applications were discussed in \cite{Fro1}. Here we want to mention another possible application which is of great interest for future development 
of some experimental and theoretical approaches to describe the actual optical spectra of various atoms and ions. The central idea of this method is simple: let us replace the 
actual (or exact) matrix elements $W_{ij}$ of the Coulomb potential (or Coulomb potential energy) by some varied non-linear parameters $w_{ij}$. Then, by using the known 
experimental data (total energies of some low-lying bound atomic states) and our formulas derived above we can approximate to relatively high accuracy a large number of atomic 
bound states (or energy levels) which belong to the same atomic term $\Bigl[ L, M, S, S_z, \pi \Bigr]$. For different atomic terms one needs to use different sets of the varied 
non-linear parameters $w_{ij}$. This simple semi-empirical approach can be used to simplify the actual analysis, classification and theoretical description of the bound state 
spectra of different atoms and ions. To emphasize the importance of this problem let us note that currently only 21 \% - 25 \% of all known experimental bound state spectra of 
multi-charged ions have ever been studied and described theoretically.


\begin{thebibliography}{01}
\bibitem{Virial} Sometimes, the operator $T = r p_{r}$ is called the `virial operator' (or, virial, for short). The unitary operator $\exp(\imath \chi T)$
is called the `virial rotation'.
\end{thebibliography}
\end{document}